# Dust and Ices in the SNR

A.Yeghikyan[1],* M. Rah[1,2], S. Shamyar[1]

[1] NAS RA V.Ambartsumian Byurakan Astrophysical Observatory (BAO), Byurakan, Armenia

[2] The Silk Road Project at the National Astronomical Observatory, Chinese Academy of Sciences, China

**Abstract**

The presence of dust in supernova remnants (SNRs) is confirmed by extensive infrared data from observatories such as Spitzer, Herschel, and JWST, alongside theoretical models of dust formation. This study explores the existence of dust and ices, particularly water ice via 62 $\mu$m in SNRs such as the Crab Nebula and N49, using observational data and preliminary modeling with Cloudy. Observations suggest that water ice may be present in IC 443 and possibly other remnants, though the 63 $\mu$m band could also indicate [OI] emission. Theoretical models indicate that water ice could survive under certain conditions in SNRs, with densities and temperatures analyzed. Further observations and refined simulations are needed to confirm these findings.

**Keywords:** Supernova Remnants, Dust, Ices, Water Ice, Crab Nebula, N49, Cloudy.

## 1. Introduction

### 1.1. Importance of Dust and Ice in Supernova Remnants

Currently, the presence of dust in supernova remnants is beyond doubt due to the large amount of observational data obtained in the IR range by the space observatories Spitzer, Herschel, JWST, etc. (Milisavljevic et al., 2024, Szalai et al., 2025) as well as many theoretical calculations on the possibility of dust particle formation under these conditions (Todini & Ferrara, 2001). Dust grains play a critical role in shaping the interstellar medium (ISM) and influencing star formation processes, making their study in SNRs essential for understanding galactic evolution.

### 1.2. Role of Supernovae as Dust Sources in the Universe

It is well known that stars with masses $M$ (init) $> 8 - 9\ M_\odot$ should explode as supernovae and form remnants in the form of expanding high-speed gas, such as the famous Crab. By the way, neutron stars are formed during the explosion at $M$ (init) $> 8\ M_\odot$ and black holes at $M$ (init) $> 20 - 40\ M_\odot$. In this case, the masses of the ejected gas range from several to tens of solar masses (Howell, 2019). Note that using the specific example of SN 1987, it was shown (Dwek & Arendt, 2015) that during the supernova explosion, the amount of dust formed was 0.45 $M_\odot$, including 0.4 $M_\odot$ - silicate and 0.05 $M_\odot$ - carbon. Huge quantities of dust observed in high-redshift galaxies (Nozawa et al., 2005) raise the question of dust origin, with supernovae proposed as dominant sources due to their rapid dust production compared to low-mass stars.

### 1.3. Challenges of Molecular and Ice Phase Survival in Energetic Environments

The problem of ices is more complex in terms of existence in such highly excited nebulae as SN remnants. However, molecules are observed quite reliably in these objects, as established starting from early observations of $H_2O$ and OH in Kes 45 (G342.1+0.1) (Sandell et al., 1983), CO, $H_2CO$ in W44 (Dickel et al., 1976) and up to later registrations in IC443 (H$_2$O, CI, $^{13}$CO, CO, HCO$^+$ (Snell et al., 2005)). The survival of ices like water ice under the intense UV radiation and high-energy particle fluxes in SNRs poses a significant challenge, necessitating detailed modeling and observational confirmation.

*ayarayeg@gmail.com, Corresponding author

## 2. Water Ice in SN Remnants – Observations

### 2.1. Previous Observational Evidence

Some SNRs exhibit a 62 $\mu$m band, potentially indicating water ice in crystalline form or the [O I] 63 $\mu$m line. Reports of water (gas and ice) in IC443, N49, W44, and Kes45 highlight the presence of molecular species in these environments. The key spectral bands include the 62 $\mu$m (crystalline H$_2$O) and 63 $\mu$m ([O I]) lines, which are critical to identifying ice signatures. Systematic searches using Spitzer archival data (Reach et al., 2006) identified IR counterparts for 96 SNRs, with follow-up observations confirming H$_2$ and CO emissions (Hewitt et al., 2009). The spectrum of N49 ( LMC) provides a prime example of these features.

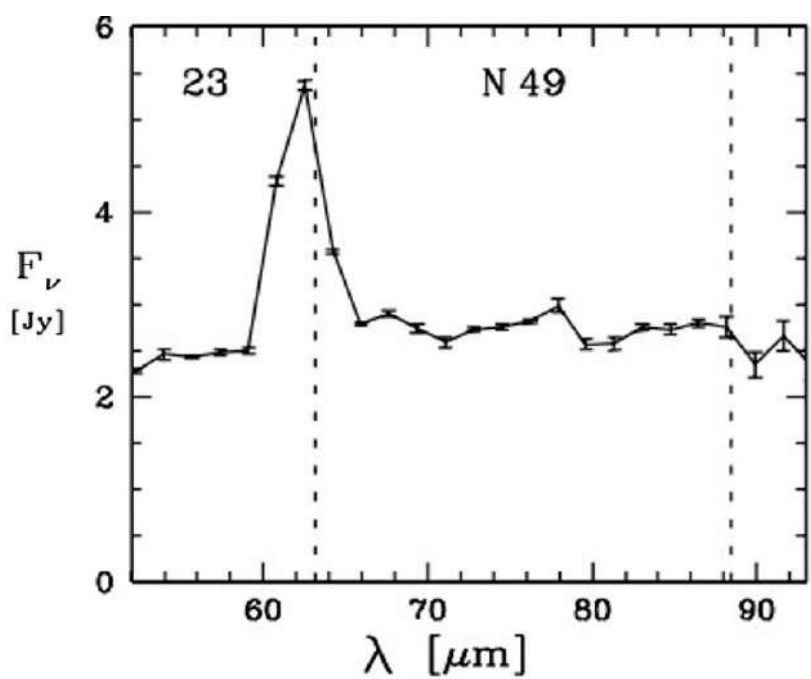

Figure 1. The profile of 62 $\mu$m band in the spectrum of N49 - SNR in LMC,(van Loon et al., 2010), about 5000 year old. It relates to the crystalline water ice emission - the lattice-mode feature at 62 $\mu$m . Dashed vertical lines show positions of 63$\mu$m [O I] and 88 $\mu$m [O III].

### 2.2. Spectral Confusion and Interpretation

Figure 1 shows the spectrum of N49 ( LMC) with dashed lines at 63 $\mu$m ([O I]) and 88 $\mu$m ([O III]) (van Loon et al., 2010). Double-humped profiles at 62 and 63 $\mu$m in other SNRs suggest both emissions (Andersen et al., 2011). There is a discussion about the potential overlap of the 62 and 63 $\mu$m bands—whether this indicates the presence of ice or the forbidden [O I] line. With dust temperatures of 30-50 K, [O I] 63 $\mu$ m is probably dominant, as crystallinity requires a temperature above 110 K. However, IC 443 shows water vapor behavior hinting at a frozen state (Snell et al., 2005), suggesting possible ice survival under specific conditions. The fitting of the dust model using SEDs (Bernard et al., 2008) indicates higher abundances of polycyclic aromatic hydrocarbons (PAHs) and very small grains (VSGs) in SNRs compared to the Milky Way.

Figure 2 illustrates densities for H$_2$, H$_2$O gas, and H$_2$O ice, calculated by Cloudy (see below), highlighting the importance of water ice.

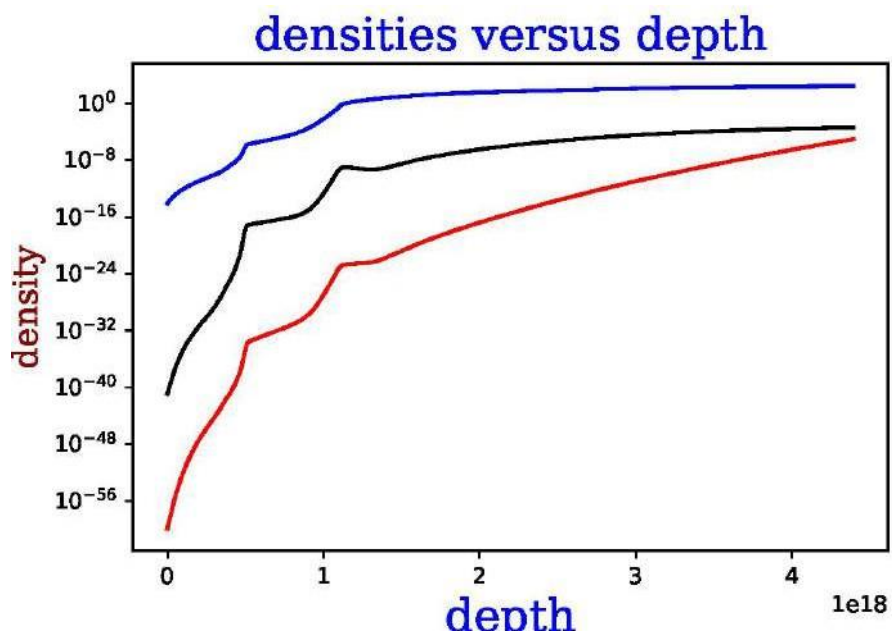

Figure 2. Distribution (depth-dependent) of molecules calculated by Cloudy: $H_2/H$(total) shown in blue, water vapor $H_2O$ (gas, cm$^{-3}$) in black, and $H_2O$ (ice, cm$^{-3}$) in red, for the Crab model with $n$(H) = 1300 cm$^{-3}$.

## 3. Theory

### 3.1. Physical Conditions in the Crab Nebula

The Crab Nebula, a composite of a pulsar wind nebula (PWN) and SNR, exhibits a complex structure with a central pulsar driving high-energy photon and particle fluxes. Pulsar wind nebulae (PWNe) are powered by the rotational energy loss of the central pulsar, creating dynamic environments rich in relativistic particles and magnetic fields (Rah et al., 2024).According to this study, pulsars are emerging and highly significant phenomena in astronomy that warrant further investigation, particularly in their role within PWNe. Studies on dust irradiation in related environments, such as planetary nebulae, provide valuable insights into ice survival under high-energy conditions (Yeghikyan, 2017). This research uses cloudy modeling and analyzes the 44 and 62 µm bands to assess the impact of proton and alpha particle irradiation on the survival of frozen water. Furthermore, UV dose calculations in molecular clouds offer a comparative framework, highlighting the radiative processing and chemical transformation of ices like $H_2O:CH_3OH:NH_3:CO$ (Yeghikyan, 2009). These clumps, with densities up to $10^4$ cm$^{-3}$, can allow ice survival against the harsh radiation environment.

### 3.2. Modeling with Cloudy

The Cloudy photoionization code was employed to simulate the physical conditions in the Crab Nebula. Calculations were performed with version (c23.01) of Cloudy (Chatzikos et al., 2023). The input parameters included a steady-state model with solar chemical composition, a Crab mode for radiation, and a radial density profile decreasing as $n(r) \propto r^{-2}$. Temperature and radius were key variables, with the simulation extending beyond $R > 10^{17}$ cm.

### 3.3. Simulation Results

Preliminary Cloudy modeling of the Crab Nebula shows water ice presence. Figure 3 depicts dust and gas temperatures, with $T_{dust} < 100$ K beyond $R > 10^{17}$ cm, suggesting non-crystalline dust. Denser clumps may exceed 100 K, akin to young stellar objects, making SNR clumps promising for ice survival (Ferland et al., 2017). The simulation estimated a water ice mass of approximately $10^{-10}$ $M_\odot$, consistent with low-abundance survival. The shock physics and the cooling mechanisms, as described in (Hollenbach & McKee, 1989), were considered to refine the model.

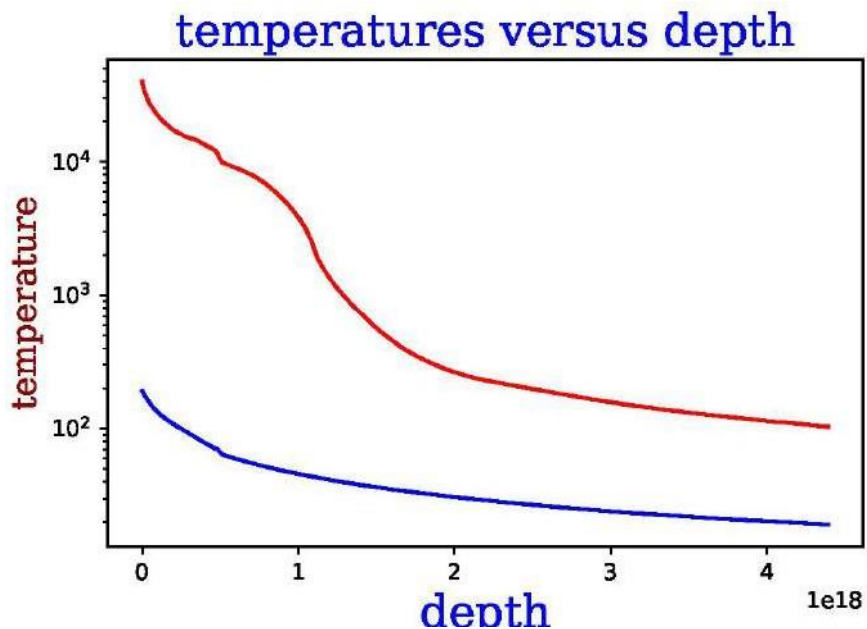

Figure 3. Distribution of gas (red) and dust (blue) temperatures calculated by Cloudy for the same Crab model. Only for R > 5E17 cm dust temperature is less than 100 K, that is ice H2O is in amorphous mode, as for crystallinity dust temperature should be above 110 K.

## 4. Discussion

The role of dense clumps in the preservation of water ice is critical, as they shield against UV radiation and high-energy particles (Draine, 2003). Compared to planetary nebulae (PNs), where ice survival is more common due to lower radiation fields, SNRs present a harsher environment, yet clumps may replicate PN-like conditions locally. Observations of dust processing in SNRs (Rho et al., 2008, 2009, 2012a,b) indicate significant grain destruction and reformation. Why only water ice persists while CO and OH ices are less evident may be related to the higher binding energy of water and lower sublimation temperature. The refinements of the dust model (Compiègne et al., 2008) and the studies of grain evolution (Temkin et al., 2010) suggest that shock interactions enhance PAH and VSG abundances.

## 5. Conclusion

Observational data suggest water ice in IC 443 and possibly other SNRs, although confirmation requires further study. Theoretical models support the survival of the ice under specific conditions, with Cloudy indicating $10^{-10}$ $M_\odot$ of water ice in the Crab without silicates. This has implications for interstellar molecular chemistry, as water ice could seed complex molecule formation. Future observations with JWST, supported by the Herschel findings (Barlow et al., 2010), are recommended to validate these findings.